\documentclass[useAMS,usenatbib]{mn2e}

\usepackage{psfig, epsf, epsfig}

\title[Star formation in almost dark galaxies]{
Formation of emission line dots and
extremely metal-deficient dwarfs from almost dark galaxies}
\author[K. Bekki]
{Kenji Bekki${}^1$\thanks{E-mail:
bekki@cyllene.uwa.edu.au} \\
${}^1$ICRAR M468
The University of Western Australia
35 Stirling Hwy, Crawley
Western Australia 6009, Australia}

\begin{document}

\date{Accepted, Received 2005 February 20; in original form }

\pagerange{\pageref{firstpage}--\pageref{lastpage}} \pubyear{2005}

\maketitle

\label{firstpage}

\begin{abstract}

Recent observations have discovered a number of extremely 
gas-rich
very faint dwarf galaxies possibly embedded in low-mass dark matter halos.
We investigate star formation histories of these gas-rich dwarf
 (``almost dark'') galaxies
both for isolated and interacting/merging cases.
We find that although star  formation rates (SFRs) are
very low  ($<10^{-5} {\rm M}_{\odot}$ yr$^{-1}$) in the simulated dwarfs in isolation for
the total halo masses ($M_{\rm h}$) of  $\rm 10^8-10^9 {\rm M}_{\odot}$,
they can be dramatically increased  to be $\sim 10^{-4}$ ${\rm M}_{\odot}$ yr$^{-1}$
when they interact or merge with other dwarfs.
These interacting faint dwarfs with central compact H~{\sc ii} regions can be 
identified as isolated emission line dots (``ELdots'') owing to
their very low surface brightness envelopes of old stars.
The remnant of these interacting and merging dwarfs can finally develop
central compact stellar systems with very low metallicities ($Z/Z_{\odot}<0.1$),
which can be identified as extremely metal-deficient (``XMD'') dwarfs.
These results imply
that although  there would exist many faint dwarfs that
can be hardly detected in the current optical observations, 
they can be detected as isolated ELdots or XMD dwarfs,
when they interact with other galaxies and their host environments.
We predict that nucleated ultra-faint dwarfs
formed from the darkest dwarf merging can be identified as low-mass globular clusters
owing to the very low surface brightness stellar envelopes.
%We suggest that  young stars possibly associated with the Smith Cloud,
%which is one of the Galactic High Velocity Clouds (HVCs).
%can be due to star formation triggered 
%in  the cloud embedded in a dark matter halo.

\end{abstract}


\section{Introduction}

Recent spectroscopic observations have discovered apparently isolated compact
H~{\sc ii} regions around galaxies (e.g., Ryan-Weber et al. 2004;
Boquien et al. 2009; Werk et al. 2010, W10; Kellar et al. 2012), 
which are often referred
to as emission line dots (``ELdots''). 
Clearly, some of these ELdots were observed within stripped H~{\sc i} gas of possibly
interacting galaxies
like NGC 1533 (Ryan-Weber et al. 2004), which suggests that such ELdots
were composed purely of new stars formed from tidal debris of galaxy interaction. 
Indeed, previous numerical simulations demonstrated that star formation is possible
in the gaseous tails stripped from gas-rich disk galaxies
(Bekki et al. 2005).
Some of the ELdots can be low- or high-redshifts background galaxies
with the redshifted  emission lines detected in the narrow filter passband
adopted for observations
(e.g., W10).

Although W10 confirmed that massive star formation observed as ELdots in the outer
parts of disk galaxies tends to be associated with interacting galaxies,
they also discovered
ELdots that were located in the far  outskirts of galactic disks with
no signs of galaxy interaction. 
Such ELdots can 
be simply local  star-forming regions of the very outer disks in gas-rich
disk galaxies.
It is also possible that  star-forming low-mass dwarf galaxies are identified as ELdots,
because their stellar envelopes are too faint to be detected in the current optical
observations.
Kellar et al. (2012) have recently discovered numerous isolated (i.e., intergalactic)
H$\alpha$ dots in their H$\alpha$ survey.
The origin of these ELdots in apparently isolated environments remains unclear.

%%%%% TABLE1
\begin{table*}
\centering
\begin{minipage}{160mm}
\caption{Description of the basic parameter values
for the representative low-mass dwarf  models. 
%The comparative models M7 and M8
%are those with no stars ($M_{\rm s}=0$; ``starless dwarfs'').
 }
\begin{tabular}{ccccccccc}
Model name
& $M_{\rm h}$ ($\times 10^9{\rm M}_{\odot}$)
& $M_{\rm s}$ ($\times 10^6{\rm M}_{\odot}$)
& $M_{\rm g}$ ($\times 10^6{\rm M}_{\odot}$)
& $f_{\rm b}$ 
& $f_{\rm g}$ 
& $r_{\rm vir}$ (kpc)
& $c$ 
& $R_{\rm s}$ (kpc) \\
M1 & 1.0 & 0.63 & 6.3 & $6.6 \times 10^{-3}$  & 10.0 & 7.7 &  20.0 & 0.55 \\
M2 & 1.0 & 0.19 & 1.9 & $2.0 \times 10^{-3}$  & 10.0 & 7.7 &  20.0 & 0.55 \\
M3 & 1.0 & 1.89 & 18.9 & $2.0 \times 10^{-2}$ &  10.0  & 7.7 &  20.0 & 0.55 \\
M4 & 1.0 & 0.063 & 6.3 & $6.0 \times 10^{-3}$ &  100.0  & 7.7 &  20.0 & 0.55 \\
M5 & 0.1 & 0.063 & 0.63 & $6.6 \times 10^{-3}$ &  10.0  &  2.5 &  25.9 & 0.18 \\
M6 & 0.1 & 0.0063 & 0.63 & $6.0 \times 10^{-3}$ &  100.0  & 2.5 &  25.9 & 0.18 \\
M7 & 1.0 & 0.0 & 0.063 & $6.0 \times 10^{-4}$ & - & 7.7 &  20.0 & 0.55 \\
M8 & 1.0 & 0.0 & 0.13 & $1.2 \times 10^{-3}$ & - & 7.7 &  20.0 & 0.55 \\
\end{tabular}
\end{minipage}
\end{table*}

Recent observations have discovered H~{\sc ii} regions within a 
compact high velocity cloud and suggested that the most likely explanation
for the object is a new very faint dwarf galaxy with a ratio of neutral hydrogen
mass to $V$ luminosity of $M_{HI}/L_{\rm V} \ge 20$ (Bellazzini et al. 2015).
Stark et al. (2015) have  discovered possible evidence for star formation
in the Smith Cloud (Stark et al. 2015) 
whereas Adams et al. (2015) have discovered an isolated gas cloud
(AGC198606) that would have a mass of $6.2\times 10^5 {\rm M}_{\odot}$
yet shows no optical counterpart.
Meyer et al. (2015) have investigated the presence or absence of  UV emission
for HI compact cloud samples from GALFA-HI and ALFALFA blind HI surveys
and identified 29 candidates with UV emission. They suggested that 
such gas-rich HI clouds with UV emission can be very low-mass dwarfs
Leo P and Leo T.

It is currently unclear 
whether there is an evolution
link between ELdots in isolated environments
and  extremely gas-rich, very faint dwarf (``almost dark'') galaxies such as those
observed as  isolated compact HVCs with or without stars.
These apparently isolated clouds can be identified
as ELdots,
if the star formation rates (SFRs)  are high enough 
(SFR$ > 10^{-5}$ ${\rm M}_{\odot}$ yr$^{-1}$) 
to produce at least one O or B type stars 
(e.g., Fig. 15 in Thilker et al. 2007).
It is therefore an interesting problem whether and how 
these extremely gas-rich  dwarfs can show 
an observationally detectable star formation activity.
Because of the lack of detailed theoretical studies of star formation 
in these dwarfs,
it remain unclear in what physical conditions 
a detectable amount of star formation can be triggered
in such  almost dark  galaxies.

The purpose of this Letter is to propose a possible evolutionary link
between  ELdots and  extremely gas-rich low-mass dwarf galaxies based on
the new results of numerical simulations of almost dark galaxies.
We particularly focus on 
interacting and merging low-mass dwarf galaxies with their dark matter halo
masses (${\rm M}_{\rm h}$) equal to or less than $10^9 {\rm M}_{\odot}$.
This is because previous simulations mainly discussed
the formation of  blue compact dwarf galaxies (BCDs; Bekki 2008)  and star formation
histories (Stierwalt et al. 2015) in 
relatively high-mass ($M_{\rm h}>10^9 {\rm M}_{\odot}$) interacting/merging dwarf galaxies.
Although recent numerical simulations have demonstrated that dwarf-dwarf merging
can cause morphological transformation of low-mass dwarf galaxies
(e.g.,Kazantzidis et al. 2011; Yozin \& Bekki 2012),
these  were unable to discuss star formation histories
of the galaxies. Thus, the present new simulations 
for $M_{\rm h} \le 10^9 {\rm M}_{\odot}$  can be valuable
for understanding the origin of possibly very low-mass dwarf galaxies
like ultra-faint dwarfs (UFDs).

\begin{figure}
\psfig{file=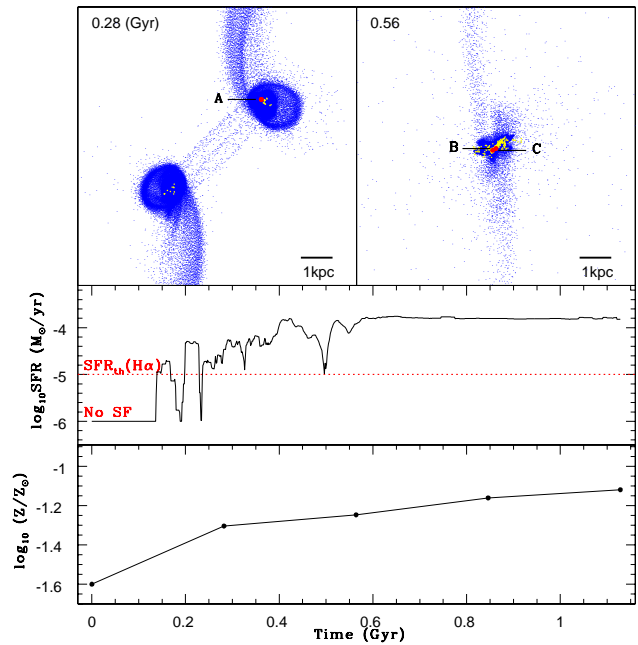,width=8.5cm}
\caption{
The distributions of gas (blue),  new stars (yellow), and H$\alpha$ regions
(big magenta circles)  at T=0.28 Gyr (top left) and
0.56 Gyr (top right),  the time evolution of the SFR (middle),
and that of the mean metallicity for new stars (bottom), in the merger model M1
with $M_{\rm h}=10^9 {\rm M}_{\odot}$.
The H$\alpha$ regions are new stars that have ages less than 1 Myr 
and are surrounded by gas particles.
For clarity, the three ELdots (H$\alpha$ regions) are indicated by
labels, A, B, and C.
The dotted red line in the middle panel
indicates the threshold SFR (SFR$_{\rm th}$(H$\alpha$)) above which the star-forming regions of a galaxy
can be detected as H$\alpha$ sources.
If the SFR of a simulated dwarf at a time step is zero,
then  it is plotted as $\log_{10} {\rm SFR}=-6$.
}
\label{Figure. 1}
\end{figure}

\section{The model}

%Although this code enables us to 
%investigate the formation and evolution of dust and molecular
%hydrogen (${\rm H}_2$) in disk galaxies (Bekki 2013, 2015),
%we do not investigate dust and ${\rm H_2}$ contents of low-mass dwarfs
%owing to the numerically  
%costly calculation of dust and  ${\rm H}_2$ formation and evolution.
%into the present simulation.

We investigate SFRs of low-mass dwarfs with $M_{\rm h} \le 10^9 M_{\odot}$ that are
in isolation or interacting/merging with other dwarfs.
In order to simulate the time evolution of
SFRs and gas contents in the dwarfs,
we use our original chemodynamical simulation code with dust physics that can be run
on GPU machines (Bekki 2013, 2015).
A dwarf galaxy  is composed of  dark matter halo,
stellar disk,  and  gaseous disk in the present study.
The total masses of dark matter halo, stellar disk, and  gas disk
are denoted as $M_{\rm h}$, $M_{\rm s}$, and $M_{\rm g}$,
respectively. The total disk mass (gas + stars)
and gas mass ratio ($M_{\rm g}/M_{\rm s}$))
are denoted as $M_{\rm d}$ and $f_{\rm g}$, respectively,  for convenience.
The baryonic mass fraction ($f_{\rm b}=M_{\rm d}/M_{\rm h}$)
in a dwarf galaxy  is assumed to be a free parameter.
We adopt the density distribution of the NFW
halo (Navarro, Frenk \& White 1996) suggested from CDM simulations
and the ``$c$-parameter''  ($c=r_{\rm vir}/r_{\rm s}$, where $r_{\rm vir}$ 
and $r_{\rm s}$ are  the virial
radius of a dark matter halo and the scale length of the halo) 
and $r_{\rm vir}$ are chosen appropriately
for a given dark halo mass ($M_{\rm h}$)
by using the $c-M_{\rm h}$ relation
predicted by recent cosmological simulations (Neto et al. 2007).

The radial ($R$) and vertical ($Z$) density profiles of the stellar disk are
assumed to be proportional to $\exp (-R/R_{0}) $ with scale
length $R_{0} = 0.2R_{\rm s}$  and to ${\rm sech}^2 (Z/Z_{0})$ with scale
length $Z_{0} = 0.04R_{\rm s}$, respectively.
The gas disk with a size  $R_{\rm g}=2R_{\rm s}$
has the radial scale length of $0.5R_{\rm g}$ and a  vertical scale lengths
of $0.02R_{\rm g}$. 
We adopt the reasonable values for $f_{\rm g}$ and $f_{\rm b}$ in the dwarf
models by using the observed correlations between $M_{\rm HI}$ and $M_{\rm s}$
and between $M_{\rm h}$ and $M_{\rm g}+M_{\rm s}$ (Papastergis et al. 2012). 
We mainly investigate the models with $M_{\rm h}=10^8 {\rm M}_{\odot}$
or $10^9 {\rm M}_{\odot}$,
$f_{\rm g}=10$ or 100, and $f_{\rm b}$ ranging from $6\times 10^{-4}$
to $2 \times 10^{-2}$.

We investigate both (i) isolated models in which dwarfs do not interact
with other galaxies and their host environments
at all and (ii) interacting/merging models in which
two {\it equal-mass} dwarfs interact or merge with each other.
The initial distance 
and the pericenter distance ($R_{\rm p}$) 
of two interacting/merging dwarfs are set to be
$10R_{\rm s}$ and $0.5R_{\rm s}$, respectively.
The orbital eccentricity ($e_{\rm p}$) is
set to be 1.2 (i.e., hyperbolic encounter) and 1 for the interacting
and merging models, respectively.
The spin of each galaxy in an interacting or  merging pair  is specified by 
angles $\theta_i$ (in units of degrees), where suffix $i$ is 
used to identify each galaxy. $\theta_i$ is 
the angle between the z axis and the vector of the angular momentum of a disk. 
We show the results of the models with $\theta_1=30$ and $\theta_2=45$.

A gas particle can be converted
into a new star if (i) the local dynamical time scale is shorter
than the sound crossing time scale (mimicking
the Jeans instability) , (ii) the local velocity
field is identified as being consistent with gravitationally collapsing
(i.e., div {\bf v}$<0$),
and (iii) the local density exceeds a threshold density for star formation ($\rho_{\rm th}$).
We mainly investigate the models with $\rho_{\rm th}=10$ atoms cm$^{-3}$.
%and the dependences of the present results on $\rho_{\rm th}$
%are briefly discussed later.
We adopt  the  Kennicutt-Schmidt law
(SFR$\propto \rho_{\rm g}^{\alpha_{\rm 1,5}}$,
where $\rho_{\rm g}$ is a gas density;  Kennicutt 1998) for star formation.
The models for chemical evolution and supernova (SN) feedback effects are 
the same as those adopted in Bekki (2013).
The initial gaseous metallicity is set to be [Fe/H]=$-1.6$ dex
in all models.

%The total numbers of particles used for dark matter, stellar disk,
%and gas disk in a dwarf-dwarf merger are 
%800000, 200000, and 100000 respectively.
The total numbers of particles used for  a merger model
is $1.1 \times 10^6$, and the mass resolution for gaseous components
is $1.2 \times 10^2 {\rm M}_{\odot}$ 
($1.1 \times 10 {\rm M}_{\odot}$) for $M_{\rm h}=10^9 {\rm M}_{\odot}$
($10^8 {\rm M}_{\odot}$).
The softening length is assumed to be the same between old stellar,
gaseous, and new stellar
particles in the present study.
The gravitational softening length for dark 
and baryonic components 
are 62 pc and 8 pc, respectively, for $M_{\rm h}=10^9 {\rm M}_{\odot}$.
These values are different in models with different $M_{\rm h}$.

We mainly present the results of six representative models (M1-M6) for each of which
isolated, interaction, and merging case are investigated.
These models show typical  behaviors in
the star formation histories of very low-mass dwarf galaxies.
The model M1 is the fiducial model, and the comparative models M7 and M8
are those in which there  are no old stars (``starless'' or ``dark galaxies'').
Table 1 summarized the model parameters for these eight models.
Thilker et al. (2007) showed that if a star-forming region 
of  a star cluster (or of a galaxy) has
its SFR  less than
a threshold SFR,
it can not be detected as a H$\alpha$ source. The threshold SFR is defined as
(Thilker et al. 2007):
\begin{equation}
{\rm SFR}_{\rm th}({\rm H}\alpha)=10^{-5} {\rm M}_{\odot} {\rm yr}^{-1}.
\end{equation}
We adopt this value in order to discuss whether a simulated galaxy
can be detected as a H$\alpha$ (star-forming) object.

\begin{figure}
\psfig{file=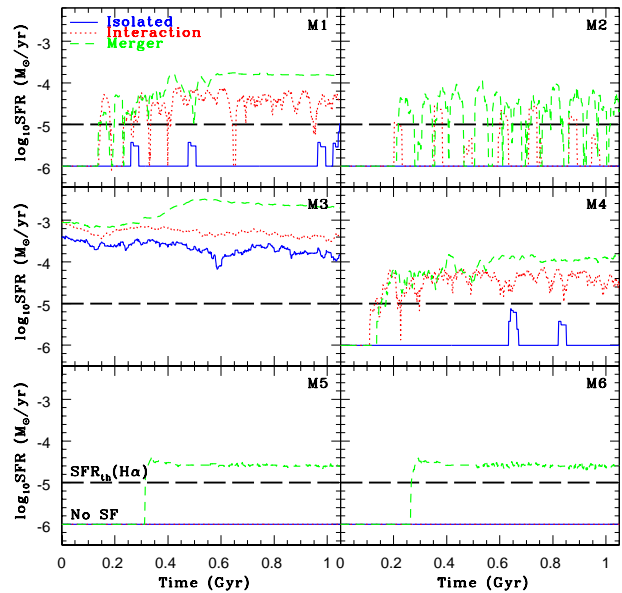,width=8.5cm}
\caption{
Star formation histories of isolated (blue solid),  interacting (red dotted),
and merging (green dashed) dwarfs in the representative 6 models (M1-M6). The thick
black long-dashed line indicates the threshold SFR for H$\alpha$ detection.
M7 shows no star formation even during interaction and merging
whereas M8 shows very low SFR only during merging (i.e., no star formation 
in isolation).
}
\label{Figure. 2}
\end{figure}

\section{Results}

Fig. 1 shows that star formation  can be ignited as dwarf-dwarf interaction
starts ($T>0.13$ Gyr) in the fiducial merger model M1 with 
$M_{\rm h}=10^9 {\rm M}_{\odot}$.
The SFR during dwarf-dwarf interaction and merging can be significantly
higher than $10^{-5} M_{\odot}$ yr$^{-1}$ 
(=SFR$_{\rm th}$(H$\alpha$)) at $T=0.28$ Gyr in this model, and 
only one of the interacting galaxies can host a compact H~{\sc ii} region.
The total stellar mass of  new stars ($M_{\rm ns}$) at $T=0.28$ Gyr is
only $3.1 \times 10^3 {\rm M}_{\odot}$, and the stellar envelopes of old stars
in the dwarfs have surface densities that are by a factor of $\sim 100$
lower than that of the Galaxy.
Furthermore, 
the mergers of the present models show that $R_{\rm 25}$ (where $B-$band surface
brightness is 25 mag arcsec$^{-2}$) is less than 100 pc.
Therefore this star-forming object in M1
is unlikely to be detected as a faint galaxy like blue
diffuse dwarfs (e.g., James et al. 2015), if its distance is
more than 20 Mpc: it could be identified as a point-source ($<1^{"}$) at such a distance.
This object with a H~{\sc ii} region can 
be identified as an isolated emission line object (ELdot).
The merger remnant at $T=0.56$ Gyr also has two neighboring H~{\sc ii} regions 
surrounded by young stars with $M_{\rm ns}=2.4 \times 10^4 {\rm M}_{\odot}$.
Although this low-luminosity remnant is also likely to be detected as an isolated
ELdot, it can be detected as an isolated gas cloud, if radio observations can detect
its cold gas mass of $M_{\rm g} \sim  10^6 {\rm M}_{\odot}$.

After dwarf-dwarf merging, the remnant can slowly build up its central compact
stellar system composed of new stars by converting
the inflowing very metal-poor gas into stars ($T>0.56$ Gyr). 
The SFR can be kept as high as $10^{-4} {\rm M}_{\odot}$ so that the final
$M_{\rm ns}$ at $T=1.1$ Gyr can be $1.0 \times 10^5 {\rm M}_{\odot}$.  As shown in Fig. 2,
the typical metallicity of  young stars 
is still low ($\log_{10} Z/Z_{\odot} \sim -1.1$)
even after merging ($T>0.56$ Gyr).
Unlike major mergers between luminous galaxies,  the low-mass dwarf-dwarf
merging does not show rapid chemical enrichment owing to the very low
star formation efficiency 
(${\rm SFR}/M_{\rm g} \sim 10^{-11}-10^{-10}$ yr$^{-1}$) and SFR.
As a result of this,   the merger remnant can have 
stellar and gaseous metallicities that are only 
slightly higher than the initial values.
These remnants can be detected  as extremely metal-poor
(i.e., 12+log(O/H)$<7.65$),
blue star-forming dwarfs when the central luminosities of young stars is high
enough.

Fig. 2 shows that SFRs in the six isolated models are either zero
or rather low ($<10^{-5} {\rm M}_{\odot}$ yr$^{-1}$), which implies that
these gas-rich, metal-poor faint dwarfs in isolation are  unlikely to be detected
by recent emission-line surveys.
The physical reason for this low SFR is that local and global dynamical
instabilities, which can trigger galaxy-wide star formation,
can be almost completely suppressed by the dominance of massive dark
matter halos in these models ($f_{\rm b}<0.007$). These dwarfs, however,  
might be detected
by  UV observations (e.g., Meyer et al. 2015), if they have $\log_{10} {\rm SFR} >
-5.5$ (e.g., Thilker et al. 2007) and if they are relatively nearby objects.
Fig. 2 also shows SFRs can be higher than $10^{-5} {\rm M}_{\odot}$ yr$^{-1}$
only during
and after dwarf-dwarf  merging in the six representative models.

Although hyperbolic tidal interaction can significantly
enhance SF in the models M1-M4 with $M_{\rm h}=10^9 {\rm M}_{\odot}$, 
it can not ignite 
SF in low-mass models M5 and M6 with $M_{\rm h}=10^8 {\rm M}_{\odot}$.
These results imply that major merging
is essential for such low-mass dwarfs to have a detectable
amount of SF.
The isolated model M3 with higher $f_{\rm b}=0.02$ can have SFRs that 
can be detected by emission whereas other models with lower $f_{\rm b}$
show no or little SF in the isolated
evolution.  This result suggests that $f_{\rm b}$  is a key parameter for 
SF histories
of low-mass dwarfs in isolation.
The remnants of low-mass dwarf-dwarf merging can continue their
low-level SF, because new gas disks with higher 
gas densities can be developed after gas-rich merging.
It is confirmed that minor dwarf-dwarf merging can show SFRs higher than
$10^{-5}$ ${\rm M}_{\odot}$ yr$^{-1}$, thought the mean SFR can be
significantly lower than major merging cases.

Fig. 3 shows that  the remnant of dwarf-dwarf merging with $M_{\rm h}=10^8 {\rm M}_{\odot}$,
$f_{\rm b}=0.006$, and $f_{\rm g}=100$  has a very compact nucleus 
with $R_{\rm 25}=51$ pc embedded by a
very diffuse stellar envelope
with $\mu_{\rm B} >30$ mag arcsec$^{-2}$ at $R>200$ pc.
The final total masses of old
and new stars are $6.3 \times 10^3 {\rm M}_{\odot}$ and 
$2.3 \times 10^4 {\rm M}_{\odot}$, respectively. The compact nucleus with the size
well less than 100pc can be classified as a low-mass globular cluster (GC) owing
to the very low surface brightness envelope of old stars, if it is detected in
photometric observations. 
Since the merger precursor galaxy in this model
corresponds to an UFD, this result suggests that
gas-rich UFDs can be transformed into low-mass ``GCs'' with stellar halos.
These ``GCs'' (which are indeed nucleated UFDs with 
star formation) can have dark matter halos and therefore can be distinguished
from normal GCs without dark matter.
Since this transformation from UFDs into low-mass GCs 
is a serendipitous discovery in the present study,
we need to explore the details of this transformation process in our
future studies.

The comparative  model M7 with no old stars
and a low $f_{\rm b}$ ($=6 \times 10^{-4}$)
does not show any SF in isolated, interacting, and merging cases
(thus not shown in Fig. 2).
The compact
gas cloud in the merger remnant embedded in a dark matter halo could be identified
as a high velocity cloud (HVC) in a luminous galaxy like the Galaxy, when it
enters inter the virial radius of the galaxy. The comparative merger model M8
with $f_{\rm b}=1.2 \times 10^{-3}$
shows low-level SF during and after major merging whereas the isolated model M8
does not show any SF.  These results for M7 and M8  imply that
there would be a threshold $f_{\rm b}$ above which star formation is re-activated
through merging in low-mass halos.

\begin{figure}
\psfig{file=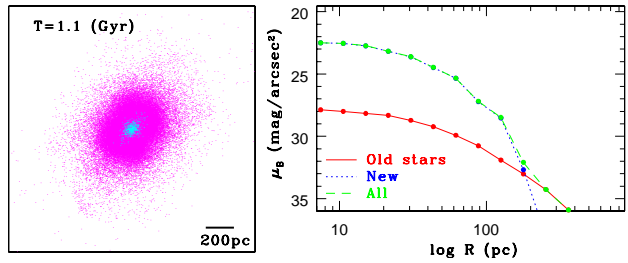,width=8.5cm}
\caption{
The distribution of old (magenta) and new (cyan) stars in the remnant
of dwarf-dwarf merging (left)  and the radial $B-$band surface brightness
profile for old (red solid), new (blue dotted), and all (green dashed) stars
in the merger  model M6 with $M_{\rm h}=10^8 {\rm M}_{\odot}$.
In estimating the $\mu_{\rm B}$ profiles, we adopted the stellar 
population synthesis code ``MILES'' (Vazdekis et al. 2010) and
assumed [Z/H]=$-1.3$ and $M/L_{\rm B}=1.7$ (old stars)
and $M/L_{\rm B}=0.17$ (new stars).
}
\label{Figure. 3}
\end{figure}

\section{Discussion and conclusions}

We have shown that extremely gas-rich,  very faint dwarf (``almost dark'') galaxies
are unlikely to have SFRs higher than $10^{-5} {\rm M}_{\odot}$ yr$^{-1}$,
if they are in isolation.
However, such almost dark galaxies can become  star-forming objects
with SFRs as high as $10^{-4} {\rm M}_{\odot}$ yr$^{-1}$ that is readily
detected in recent H$\alpha$ emission surveys (e.g., W10),
when they interact or  merge with other low-mass galaxies: These
objects with very low-level SF
transformed from almost dark galaxies might be dubbed as 
``cosmic fireflies''. 
The star-forming regions 
with SFRs much lower than $0.1-1 {\rm M}_{\odot}$ yr$^{-1}$
are compact surrounded by very low surface brightness
envelopes of old stars that can be hardly detected by optical observations.
Therefore,  such interacting/merging  dwarfs with detectable SF 
are likely to be identified as isolated
ELdots rather than starbursting BCDs.

After dwarf-dwarf interaction and merging,  star formation can continue to occur
at low level (SFR$\sim 10^{-4} {\rm M}_{\odot}$  yr$^{-1}$)
 in the star-forming low-mass dwarfs. 
The total mass of new stars ($M_{\rm ns}$) in the central compact component
of the dwarfs can be large 
($M_{\rm ns} > 10^5 {\rm M}_{\odot}$)
so that the central components can be detected as low-luminosity compact blue dwarfs.
However, the mean stellar  metallicities of the dwarfs are still
rather low ($Z<0.1 Z_{\odot}$). Therefore, such dwarfs are likely to be classified as 
XMD galaxies, if they are detected by optical and spectroscopic observations.
The present study thus suggests that
some of the observed XMD galaxies (e.g., Ekta et al. 2010;
Skillman et al. 2013; James et al. 2015)
can be the remnants of dwarf-dwarf interaction and merging.
Given that a number of observational studies have recently started to search for
very faint optical counterparts of compact HVCs (e.g., Adams et al. 2015;
Janowiecki et al. 2015;
Sand et al. 2015),
these results  provide the following important implications on the origins
of ELdots,  compact HVCs, and XMD dwarfs.

First is that there could be two different types of ELdots with low and high
metallicities.
The present study predicts (i) that ELdots formed from interacting or
merging low-luminosity dwarf galaxies show rather low metallicity 
($Z<0.1 Z_{\odot}$) and (ii) that
they might be surrounded by very low surface brightness old stellar 
components. Since ELdots formed from gas stripped from luminous disk 
galaxies (like NGC 1533) or in the very outer part of galactic gas disks
are unlikely to show very low metallicities, the metallicities of ELdots
might be a key to discriminate between different formation scenarios of
ELdots. Also, very deep optical imaging of ELdots is doubtlessly worthwhile,
because it would be able to reveal the very low surface-brightness components
of low-mass dwarf galaxies.

Second is that interacting and merging low-mass dwarfs can be identified
as compact HVCs with star formation. Recently, Stark et al. (2015) have 
revealed that there is an excess of OB stars along the line of sight
to the Smith Cloud, which is one of the massive HVCs in the Galaxy. 
They suggested that the HVC has been forming stars since it passed
through the Galaxy about 70 Myr ago.
The present study suggests that the cloud is a very gas-rich low-mass dwarf
galaxy with its star formation highly enhanced by the interaction with the Galaxy.
It is, however, unclear why such a low-mass dwarfs can retain 
much gas now, since ram pressure stripping by the Galactic halo gas
can remove all  gas from such low-mass dwarf galaxies (Yozin \& Bekki 2015).

Third is that some of the observed compact HVCs in the Galaxy
and M31 (e.g., Westmeier et al. 2005) might be the remnants of ancient
mergers between low-mass halos:
Some compact HVCs are embedded in low-mass dark matter halos.
The compact HVCs embedded in dark matter halo might be less 
susceptible to the ram pressure stripping by their host galaxies
so that the central regions can survive from such stripping
processes and thus can be identified as compact HVCs.
As demonstrated in the present study,  star formation is unlikely
in dwarf-dwarf mergers with very low $M_{\rm h}$ and $f_{\rm b}$, 
though this  might be due largely to the adopted simulation code
with no efficient cooling below $T=100$K.
No/little star formation can prevent
cold gas from thermal evaporation  of energetic supernovae 
so that the merger remnants can be still observed as compact gas clouds.

Fourth is that the observed recent star formation in nearby very faint
dwarf galaxies like Leo T (e.g., Ryan-Weber et al. 2008) 
and Leo P (e.g., Skillman et al. 2013) can be due
to their  recent interaction and merging with other low-mass faint dwarf
galaxies.
Fifth  is that there can be numerous very faint dwarf galaxies (like
UFDs) with
very low-level star formation (SFRs well less than $10^{-5}$ ${\rm M}_{\odot}$ yr$^{-1}$)
undetected in the present H$\alpha$  observations.
Gas-rich very faint dwarf galaxies located outside the Local Group
might not be detected  by current  HI surveys either,
if their gas masses are less than $10^5 {\rm M}_{\odot}$.
These almost dark dwarf galaxies will be able to be detected by
future wide-field,  ultra-deep HI observations by the Square Kilometre Array
(SKA), if they really exit.

The present study suggests that there is an evolutionary link between
almost dark galaxies,
isolated ELdots, and XMD dwarfs. 
It should be noted here, however, that the present study assumed 
low-mass dark matter halos hosting very gas-rich dwarfs.
Although  a number of ongoing observational projects  aim at detecting very gas-rich,
low-mass dwarf candidates (e.g., Meyer et al. 2015),
it is currently  unclear whether 
there really exist numerous  almost dark  dwarfs in the universe:
The observed almost dark galaxies could be  just a rare class of dwarfs.
If such galaxies are a major population in low-mass halos,  
then the present results imply
that there could be much more isolated ELdots and XMD dwarfs
at higher redsfhits,  when  interaction and merging between low-mass halos
is highly likely
to  be 
much more frequent.

\section{Acknowledgment}
I (Kenji Bekki; KB) am   grateful to the referee  for  constructive and
useful comments that improved this paper.
%Numerical simulations  reported here were carried out on
%the three GPU clusters,  Pleiades, Fornax,
%and gSTAR kindly made available by International Center for radio astronomy research(ICRAR) at  
%The University of Western Australia,
%iVEC,  and the Center for Astrophysics and Supercomputing
%in the Swinburne University, respectively.
%This research was supported by 
%resources (gSTAR) awarded under the Astronomy Australia Ltd's
%ASTAC scheme.
%on Swinburne with support from the 
%Australian government. 
%gSTAR is funded by Swin
%burne and the Australian Government's
%Education Investment Fund.
%KB acknowledges the financial support of the Australian Research Council
%throughout the course of this work.


\begin{thebibliography}{}

\bibitem[)]{}
Adams, E. A., 2015, A\&A, 573, L3

\bibitem[)]{}
Bekki, K., 2008, MNRAS, 388, L10

\bibitem[)]{}
Bekki, K., 2013, MNRAS, 432, 2298 

\bibitem[)]{}
Bekki, K., 2015, ApJ, 799, 166

\bibitem[]{}
Bekki, K., Koribalski, B. S., Ryder, S. D., Couch, W. J., 2015, MNRAS, 357, L21

\bibitem[)]{}
Bellazzini, M. et al., 2015, ApJ, 800, L15

\bibitem[]{}
Boquien, M., et al. 2009,AJ, 137, 4561

\bibitem[]{}
Ekta, B., Chengalur, J. N., 2010, MNRAS, 406, 1238

\bibitem[]{}
James, B. L., Koposov, S., Stark, D. P., Belokurov, V., Pettini, M.,
Olszewski, E. W., 2015, MNRAS, 448, 2687

\bibitem[]{}
Janowiecki, S., et al. 2015, ApJ, 801, 96

\bibitem[]{}
Kazantzidis, S., Lokas, E. L., Mayer, L.,  Knebe, A.,  Klimentowski, J.,
2011, ApJL, 740, 24

\bibitem[]{}
Kellar, J. A., Salzer, J. J., Wegner, G., Gronwall, C.,  Williams, A.,
2012, AJ, 143, 145

\bibitem[]{}
Kennicutt, R. C., Jr., 1998, ApJ, 498, 541

\bibitem[)]{}
Meyer, J. D., et al., 2015, preprint (arXiv:1506.00044)

\bibitem[]{}
Navarro, J. F., Frenk, C. S.,  White, S. D. M.,
1996, ApJ, 462, 563 (NFW)

\bibitem[]{}
Neto, A. F., 2007, MNRAS, 381, 1450

\bibitem[]{}
Papastergis, E., Cattaneo, A., Huang, S., Giovanelli, R.,  Haynes, M. P.,
2012, ApJ, 759, 138

\bibitem[]{}
Ryan-Weber, E. V., et al. 2004, AJ, 127, 1431

\bibitem[]{}
Ryan-Weber, E. V., et al. 2008, MNRAS,  395, 1476

\bibitem[]{}
Sand, D. J., et al. 2015, ApJ, 806, 95

%\bibitem[]{}
%Schmidt, M., 1959, ApJ, 129, 243

\bibitem[]{}
Skillman, E. D., et al., 2013, AJ, 146, 3

\bibitem[]{}
Stierwalt, S., Besla, G., Patton, D., Johnson, K., Kallivayalil, N.,
Putman, M., Privon, G., Ross, G., 2015, ApJ, 805, 2

\bibitem[]{}
Stark, D. V., Baker, A. D., Kannappan, S. J., 2015, MNRAS, 446, 1855

\bibitem[]{}
Thilker, D., et al., 2007, ApJS, 173, 538

\bibitem[]{}
Yozin, C., Bekki, K., 2012, ApJL, 756, 18

\bibitem[]{}
Yozin, C., Bekki, K., 2015, submitted to MNRAS

\bibitem[]{}
Vazdekis, A., et al. 2010, MNRAS, 404, 1639

\bibitem[]{}
Werk, J. K., et al. 2010, AJ, 139, 279

\bibitem[]{}
Westmeier, T., Braun, R., Thilker, D., 
2005, A\&A, 436, 101
\end{thebibliography}
\end{document}